\documentclass[conference]{IEEEtran}

\usepackage{xcolor}
\usepackage{graphicx}
\usepackage{amsmath}
\usepackage{amssymb}
\usepackage{amsthm}
\usepackage{longtable}
\usepackage{wrapfig}
\usepackage{rotating}
\usepackage[normalem]{ulem}
\usepackage{capt-of}
\usepackage{parskip}
\usepackage{hyperref}
\usepackage{cleveref}
\usepackage[ruled, linesnumbered]{algorithm2e}
\usepackage{multirow}
\usepackage{cite}
\author{
  \IEEEauthorblockN{Siddhartha Kumar\IEEEauthorrefmark{1}, Mohammad Hossein Moghaddam\IEEEauthorrefmark{1}, Andreas Wolfgang\IEEEauthorrefmark{2}}, and Tommy Svensson\IEEEauthorrefmark{4}
  \IEEEauthorblockA{\IEEEauthorrefmark{1} Qamcom Research and Technology AB, Gothenburg, Sweden  \\
  \IEEEauthorrefmark{2} Catalytic Technology AB, Gothenburg, Sweden
  \IEEEauthorrefmark{4} Chalmers University of Technology, Gothenburg, Sweden\\
  Email: siddhartha.kumar@qamcom.se, mh.moghaddam@qamcom.se, andreas.wolfgang@catalytic.se,\\ tommy.svensson@chalmers.se }

\thanks{Part of  this work has been  funded from the European  Union’s Horizon
  2020 research  and innovation programme  under grant agreement  No 101015956
  Hexa-X,  and  Horizon  Europe  research and  innovation
  programme under the  Smart Networks and Services Joint  Undertaking (SNS JU)
  under  grant agreement  No 101095759  Hexa-X-II. The  authors would  like to
  acknowledge the contributions of their  colleagues in Hexa-X and Hexa-X-II.}
}

\title{Path Assignment in Mesh Networks at the Edge of Wireless Networks}

\hypersetup{
 pdfauthor={Andreas Wolfgang, Siddhartha Kumar, and Mohammad Hossein Moghaddam},
 pdftitle={Path Assignment in Mesh Networks at the Edge of Wireless Networks},
 pdfkeywords={},
 pdfsubject={},
 pdfcreator={Emacs 29.1}, 
 pdflang={English}}

\let\symup\mathsf
\let\symbf\mathbf
\let\symcal\mathcal

\theoremstyle{definition}
\newtheorem{example}{Example}
\newtheorem{theorem}{Theorem}
\IEEEoverridecommandlockouts
\begin{document}

\maketitle
\begin{abstract}
  We consider a mesh  network at the edge of a  wireless network that connects
  users to the core network via  multiple base stations. For this scenario, we
  present  a  novel  tree-search-based  algorithm  that  strives  to  identify
  effective communication path to the core network for each user by maximizing
  the signal-to-noise-plus-interference  ratio (SNIR)  along the  chosen path.
  We  show that,  for  three mesh  networks of  varying  sizes, our  algorithm
  selects paths with  minimum SNIR values that  are 3 dB to 18  dB higher than
  those obtained through an algorithm  that disregards interference within the
  network, 16 dB to  20 dB higher than those chosen randomly  by a random path
  selection  algorithm, and  0.5 dB  to  7 dB  higher compared  to a  recently
  introduced  genetic algorithm  (GA).  Furthermore,  we demonstrate  that our
  algorithm has lower computational complexity  compared to the GA in networks
  where its performance is within 2 dB of ours.
\end{abstract}

\section{Introduction}
\label{sec:orgc115430}
Fixed  wireless  access  (FWA)  networks  play a  crucial  role  in  providing
high-speed, fiber-like  internet connectivity  without the need  for extensive
physical infrastructure. To meet the growing demands for higher throughput and
low latency,  FWA networks require  increased signal dimensions, which  can be
achieved  either by  deploying  a larger  number of  antennas  to exploit  the
spatial  dimension or  by increasing  the signal  bandwidth at  higher carrier
frequencies~\cite{9390169}. 
The key performance indicators in FWA networks, including throughput, latency,
and reliability, are heavily influenced by efficient spectrum usage and robust
signal propagation  ~\cite{hilt2023threshold}. However, as  higher frequencies
suffer from increased  path loss, interference, and  blockages, optimizing the
communication path and  managing interference are critical  to maintaining the
desired quality of service.

The main challenge at higher carrier  frequencies is the increased signal path
loss, which  makes it more difficult  to generate the output  power needed for
long-distance  wireless  links~\cite{gatech23}.   To  effectively  use  higher
frequencies, both  ends of a communication  link must be positioned  closer to
each other.  For applications where  coverage is a priority,  this requirement
leads to network densification. As a result, densification drives the need for
efficient  network roll-out  and  backhaul solutions  to  keep deployment  and
ownership costs  low. Various  solutions to address  the problem  of last-mile
backhaul  have  been  explored  and   incorporated  into  standards,  such  as
self-backhaul~\cite{TS38174}. Self-backhaul  offers the advantage  of hardware
reuse between access  and backhaul functions, but it presents  the drawback of
backhaul and access competing for the same frequency resources.

In this  paper, we investigate  a network  topology where backhaul  and access
operate on  different carrier frequencies.  The underlying assumption  is that
the access  link operates  at a  lower carrier  frequency, while  the backhaul
operates   at  a   higher  carrier   frequency  where   larger  bandwidth   is
available.  For the  backhaul network,  we consider  a mesh  network topology,
which  is  particularly attractive  from  a  network roll-out  and  deployment
perspective.  The network  can be  easily expanded  by adding  new micro  base
stations (BSs)  without the need  for additional fiber  or macro BS  sites for
backhaul. In the proposed concept, each BS can function as both an access node
for users  and a forwarding node  in the backhaul network.  This system design
simplifies roll-out and densification  but introduces additional complexity at
the network  layer, where optimal  routing paths through  the mesh need  to be
determined.

Routing  in wired  mesh networks  has  been extensively  studied. However,  in
contrast  to  wired  mesh  networks,  optimizing  wireless  networks  is  more
challenging due  to the interdependence  of link costs caused  by interference
between different backhaul links. Routing choices directly affect interference
levels, and  interference, in turn,  influences routing decisions.  This paper
addresses this challenge by introducing a tree-search-based algorithm that determines
the optimal  route for each  user in a mesh  network while accounting  for the
interference  within   the  network.  Moreover,  the   algorithm  can  balance
performance and  scalability, depending  on the  specific requirements  of the
network operator.

Significant  research  efforts  have  been dedicated  to  optimizing  wireless
networks~\cite{Yua18,    Hu17,     Vu19,    8876705,    madapatha2021topology,
  gupta2022system}.  While  \cite{Yua18}  and \cite{Hu17}  utilize  algorithms
based  on numerical  optimization principles,  \cite{Vu19} and  \cite{8876705}
present  innovative  approaches  leveraging machine  learning  techniques.  In
\cite{madapatha2021topology},  the authors  address  the  problem of  topology
optimization and routing  for integrated access and backhaul  networks using a
genetic  algorithm.   In  \cite{gupta2022system},  topology   optimization  is
achieved by  considering latency  gains and  the maximum number  of hops  in a
mmWave, full-duplex backhaul network.

In contrast  to these works,  the algorithm in this paper distinguishes  itself  by carefully  accounting  for  network
interference, unlike  \cite{Yua18} and \cite{Vu19}. Unlike  \cite{Hu17}, where
interference  mitigation primarily  focuses on  building reflections  in urban
scenarios, our  algorithm considers interference impacts  during communication
between  base  stations  (BSs).  Additionally,  our  algorithm  is  versatile,
accommodating networks with  any number of BSs connected to  the core network,
whereas \cite{Vu19},  \cite{8876705}, and \cite{gupta2022system}  focus solely
on scenarios involving  a single core BS.  Furthermore, unlike \cite{8876705},
we do not assume that the BSs communicate at fixed times.

\textbf{Notation.} We denote a finite
  integer intervals \(\{0,\ldots,b-1\}\), and \(\{a,\dots,b-1\}\),
  \(a,b\in\mathbb{Z}\) as \([b]\) and \([a,b]\) respectively.




\section{System Model}
\label{sec:org62bb8d5}

\begin{center}
\begin{figure}[t]
\centering
\includegraphics[height=0.3\textwidth]{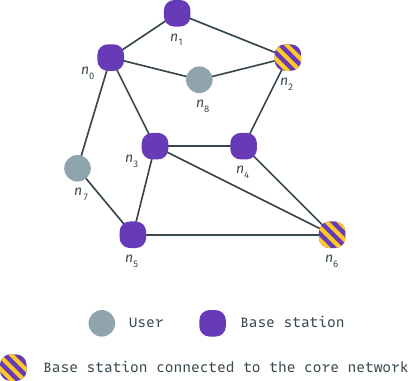}
\caption{\label{fig:system_model}Mesh network with \(U=2\) users, \(B=7\)
  BSs and \(C=2\) core BSs.}
\vskip -1em
\end{figure}
\end{center}

We consider an FWA mesh network wherein  \(U\) users, \(u_{0}, u_{1},\dots, u_{U-1}\), are
served by a network of  \(B\) base stations (BSs), \(b_0,b_1,\ldots,b_{B-1}\),
(i.e.,  the BSs and the users are connected to  each other via
established microwave links), at the edge of a  wireless network.  Of the \(B\) BSs
in the network, there are \(C\) BSs that are connected to the core
network. We refer to them as the \emph{core BSs}. Each user is connected to two BSs that  are nearest to it, and the goal of  the network is to ensure each user  is connected to the  core network via paths  in the network such that  they have  the highest  connection throughputs/lowest  latency. An example   of  a  mesh   network   that   we   consider   here   is   shown   in Fig. \ref{fig:system_model}.  
From hereon, where convenient, we would refer  to the BSs, and users as simply
the nodes of the mesh network where the nodes \(n_i,i\in[B]\) refer to the BSs
while the nodes \(n_i, i\in[B,B+U]\) are referred to the users.

For a  user \(n_i\)  and core BS \(n_j\) we  define a  valid path,
\(\rho\) if  \(n_i\) can reach  \(n_j\) by hopping  at most
\(h_{\symup{max}}\) unique  nodes (including  \(n_j\) but  excluding \(n_i\))
through established  links.  We  mathematically denote \(\rho\)  as a  set of
tuples \(\left( i_1,i_2  \right)\) of length two  representing an established
link  between  the  transmitter  node   \(n_{i_1}\)  and  the  receiver  node
\(n_{i_2}\). Naturally,  the cardinality of the  set is \(h\), the  number of
nodes hopped  from \(n_i\) to \(n_j\).   This allows us to  define a sub-path
\(\sigma\)   of    \(\rho\)   as    a   subset    of   the    latter,   i.e.,
\(\sigma\subset\rho\).

\vspace{-2em}
\subsection{Link Description}
\label{sec:orgff5c79c}

We characterize the configuration of the \(U\) users and \(B\) the
  BSs  by an  adjacency matrix, \(\symbf{A}=[a_{ij}]\)  of dimension
  \((B+U)  \times  (B+U)\).   Then,  the  neighbors  of   the  \(i\)-th  node,
  \(\symcal{N}_i\)  is  simply   the  nodes  whose  index  lies   in  the  set
  \(\symup{supp}\{\symbf{a}_i\}\backslash i\), where  \(\symbf{a}_{i}\) is the
  \(i\)-th         row         of          \(\symbf         A\),         i.e.,
  \(\symcal{N}_i=\symup{supp\{\symbf{a}_i\}}\backslash i.\)

We consider  the users to be  fixed and have direct  line of sight
  (LoS) with two nearest BSs.  Furthermore, the user and the BS have dedicated
  time/frequency resources  to achieve interference free  communication. Thus,
  they have perfect links.  On the contrary,  BSs that serve as a link between
  users and the core network share  the same time/frequency resource, and they
  might not have a direct LoS with each other leading to them having imperfect
  links.   We characterize  the environment  around these  links by  rain fade
  margin,  \(\symup{FM}_{\symup{rain}}\).  Additionally,  over long  distance,
  the communication signal across these  links suffer from signal attenuation,
  \(\symup{Att}_{\symup{O}_2}\) and noise.
  Each base station has a single transmitter and a single receiver directional
  antenna, which they use to simultaneously communicate with their neighbors.

\textbf{Received  Power.}  Let  \(P^{\symup{tx}}_j\)  be the  transmit  power  of  the
\(j\)-th BS \(n_j\),  \(j\in[B]\), then  the recieved
power  \(P^{\symup{rx}}_{ij}\),  of  the   \(i\)-th  BS,  \(n_i\),
\(i\in\symcal{N}_j\), in dBm, is given by

\begin{equation*}
\begin{split}
  P^{\symup{rx}}_{ij}=P^{\symup{tx}}_j
  +G_i(0)&-20\log_{10}\frac{4\pi fd_{ij}}{c}
  -d_{ij}\symup{FM}_{\symup{rain}}\\
  &-d_{ij}\symup{Att}_{\symup{O}_2}+G_j(0),
\end{split}
\end{equation*}

where,  \(G_i(\theta) (G_j(\theta)\)) is  the  antenna gain  in  dBm for  the
\(i\)-th (\(j\)-th)  BS  that is  a function  of the  incidence angle
\(\theta\),  \(c\)  is  the  speed  of   light  in  meters  per  second,  and
\(d_{{ij}}\)  is the  distance between  the  \(i\)-th and  \(j\)-th nodes  in
meters.
 For established links, we assume \(\theta=0\), that is the transmitter and receiver are directed towards each other.

\begin{center}
\begin{figure}[t]
\centering
\includegraphics[width=0.3\columnwidth]{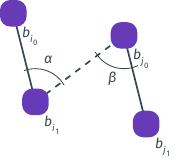}
\caption{\label{fig:Ex_interference}A sample interference configuration that occurs at \(b_{i_1}\).}
\end{figure}
\end{center}

\textbf{Interference.}  In our  model,  we consider  that the  BSs
  share the same communication resources. Therefore, communication between one
  pair of BSs  in the network interferes with another  pair. Additionally, the
  total  interference faced  by a  BS  is the  sum of  interferences that  are
  generated by all active BS pairs in the network.
Assume four BSs,  \(b_{i_0}, b_{i_1}, b_{j_0},
b_{j_1}\), where  \(b_{i_0}\) and \(b_{j_0}\)  have an established  link,
are transmitting  to \(b_{i_1}\) and  \(b_{j_1}\), respectively, as  shown in
Fig. \ref{fig:Ex_interference}.   Then, the interference power  at \(b_{i_1}\) from
\(b_{j_0}\) is

\begin{equation*}
\begin{split}
  P^{\symup{in}}_{i_1i_0j_0j_1}=P^{\symup{tx}}_{j_0}
  +G_{i_1}(\alpha)&-20\log_{10}\frac{4\pi fd_{i_1j_0}}{c}
  -d_{i_1j_0}\symup{FM}_{\symup{rain}}\\
  &-d_{i_1j_0}\symup{Att}_{\symup{O}_2}+G_{j_0}(\beta),
\end{split}
\end{equation*}

where,  \(\alpha\) is  the angle  of incidence  of the  received signal  from
\(b_{i_0}\) and  \(b_{j_0}\) at  \(b_{i_1}\), and \(\beta\)  is the  angle of
incidence of the transmitted signal  from \(b_{j_0}\) towards \(b_{i_1}\) and
\(b_{j_1}\).
  One can  then formulate  the total  interference power  faced by
  \(b_{i_1}\)      when      communicating       with      \(b_{i_0}\)      as
  \(P^{\symup{tot},\symup{in}}_{i_1}=\sum_{(j_0,j_1)\in\mathcal{I}}
  P^{\symup{in}}_{i_1i_0j_0j_1}\), where \(\mathcal I\) is a set of index
  pairs that represent BS pairs that are sharing the same resources as
  \(b_{i_0}\) and \(b_{i_1}\).

\vspace{-2em}
\subsection{Objective}
\label{sec:org4f4ad16}

As stated earlier, our objective is to  ensure that all users that are served
by  the BSs in  the network,  are reliably  connected to  the core
network,  while maintaining  highest possible  througputs.  To  this end,  we
assign a cost to each established link  in the mesh network, compute the cost
of each path from user to the  core network, and finally for each user choose
the path to the core network that  has the least cost amongst the plethora of
paths in the network.

Consider  two  nodes  \(n_i\)  and  \(n_j\), in  the  network  that  have  an
established   link   between  them,   where   \(n_j\)   is  transmitting   to
\(n_i\). Then, we formulate the link cost, \(C_{ij}\) as the SNIR at the node
\(n_i\), i.e.,

\begin{equation}
\label{eq:3}
C_{ij} =
    \frac{P^{\symup{rx}}_{ij}}{P^{\symup{n}}_i+P^{\symup{tot},\symup{in}}_i},
\end{equation}

where   \(P^{\symup{n}}_i\)   is   the    noise   power   at   \(n_i\),   and
\(P^{\symup{tot},\symup{in}}_i\)  is the  total interference  power that  the
node  \(n_i\)  encounters.   Do  note that,  by  definition,  \(C_{ij}\not=
C_{ji}\).

Given the link  cost and \(N_i\) valid paths from user \(n_i\) to a core BS, we can then  define the cost of  a path, \(\rho_{ij}\),
\(j\in[N_i]\)  to the user, \(\hat{C}_{ij}\)  as the  maximum link  cost
(i.e., minimizing the SNIR) amongst all links in the path. In other words,

\begin{equation}
\label{eq:4}
\hat{C}_{ij}=\min_{(i',j')\,\in\, \rho_{ij} } C_{i'j'}.
\end{equation}

This makes sense as in most real  world scenarios, the cost of transmission is
dictated by the  slowest link within the path. Finally,  the best path for
the user is simply the \(\hat{j}\)-th path, \(\rho_{i \hat{j}}\), where

\begin{equation}
\label{eq:5}
\hat{j}= \underset{j\,\in\,[N_i]}{\mathrm{argmax}}\; \hat{C}_{ij}.
\end{equation}

That is to  say that the optimal path  is the one which has  the highest path
SNIR. The min-max  problem is Eqs. (\ref{eq:4})  and (\ref{eq:5}) is not as  simple as it
seems.   This  is because  the  total  interference power for  node \(n_i\),  \(P^{\symup{tot},
\symup{in}}_{i}\), in  Eq. (\ref{eq:3}),  depends upon the  chosen paths  of users
 \(n_{j'}\,\forall \,j'\in[B,B+U]\backslash  i'\), where \(n_i\) is part of
a path between user \(n_{i'}\) and a core BS. 

\section{Tree Search Algorithm}
\label{sec:org6484d8d}

We now  present a  novel tree  search algorithm  that strives to find optimal
paths for the users in the mesh network. These paths are optimal in the sense
that they  are the best paths  that provide  the highest throughput  between the
users and the  core network.  In this  section, we will briefly describe our proposed algorithm.

\begin{center}
\begin{figure}[t]
\centering
\includegraphics[height=0.3\textwidth]{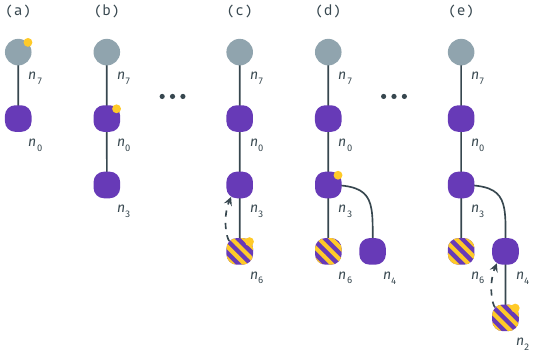}
\caption{\label{fig:Ex_tree_construction}Initial steps in the construction of the Tree, \(T_0\). At each step the node that has  a yellow circle in the top right corner represents position of the algorithm. The subsequent node represents the node that the algorithm chooses to explore.}
\end{figure}
\end{center}
\vskip -2em
For each user \(u_i\) we create a \emph{tree-like} structure \(T_i\) with \(u_i\) as
its root. The \(T_i\) tree is constructed as follows.  We start with the root
node, \(u_i\)  in the mesh network,  and sequentially explore all  nodes in a
\emph{depth-first} fashion and  add them to the tree to  form a branch. Furthermore,
we are  traversing only those  nodes that appear  only once when  forming the
branch.  We backtrack our  search once we have either explored  a node in the
mesh that represent the core BS, or when we have hopped the maximum
number of  hops permissible,  \(h_{\symup{max}}\) to  this node.   This node,
then, becomes the leaf of the  thus constructed branch\footnote{A branch refers  to a path between the root node and
the leaf nodes.} in the tree.  We
backtrack by one node (i.e., one hop) in the mesh, and then attempt to create a
new branch  by continuing the search  in the depth-first fashion.   Again, we
make sure  that we visit  only those nodes that  have not appeared  before in
this branch.  We repeat this process until  we have explored all nodes in the
mesh.

\begin{center}
\begin{figure}[t]
\centering
\includegraphics[height=0.3\textwidth]{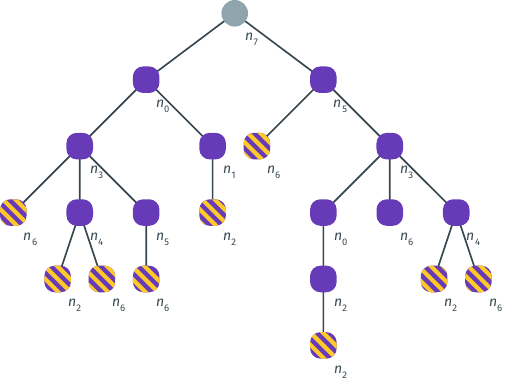}
\caption{\label{fig:Ex_tree}Tree, \(T_0\) corresponding to the user \(u_0\) in the mesh network shown in Fig. \ref{fig:system_model}.}
\end{figure}
\end{center}

\begin{example}
  We illustrate
  the  steps to  create  a  tree \(T_0\) in Fig.~\ref{fig:Ex_tree}  for  \(u_0\)  (i.e., \(n_7\))  using
  Fig. \ref{fig:Ex_tree_construction} for guidance.  Starting  with \(n_7\), we explore its
  neighbors \(\mathcal{N}_{7}=\{n_0,n_5\}\) in a  depth first fashion. Assume
  the algorithm picks \(n_0\)  to explore (\Cref{fig:Ex_tree_construction}(a)). It proceeds
  through   the  nodes,   disregarding   those  that   have  previously   been
  explored. This  process continues until  the algorithm reaches a  core node,
  say \(n_6\)  (see \Cref{fig:Ex_tree_construction}(c)),  thus completing  a branch  in the
  tree. At this  point, we backtrack to \(n_3\) and  choose an unvisited node,
  like \(n_4\),  and repeating the process until  we reach another core  node, say
  \(n_2\). This is illustrated in steps (d) and (e) of \Cref{fig:Ex_tree_construction}. The
  two  steps of  exploration  and backtracking  continue  until the  algorithm
  arrives back to  the root node \(n_7\).  At this point, it  has explored all
  nodes in the network.

\end{example}

Once the trees are created, we then  begin searching for the  path with lowest cost for
each  user.  For  the  \(i\)-th user,  we  define a  set  of dependent  trees
\(\left\{ T_{i'} \right\}_{i'\in[U] \backslash  i}\). Each of these dependent
trees have  \(N_{i'}\) valid paths,  i.e., the paths  that end with  the core
node   and  have   length  at most   \(h_{\symup{max}}\).   Furthermore,   let
\(\symcal{P}_{i'}\) be the  set of all valid paths in  \(T_{i'}\).  Then, for
each  user   \(i\),  and   the  \(j\)-th   combination  of   dependent  paths
\((\hat{\rho}_{i'j})_{i'\in[U] \backslash  i} \in  \prod_{i'\in[U] \backslash
i} \symcal{P}_{i'}\)  we find (whose  description is presented in \Cref{sec:org6853c8d}) the best path, \(\hat{\rho}_{ij}\)  in \(T_i\), along with its cost,
\(\hat{C}_{ij}\), and costs of the dependent paths \(\left\{ \hat{\rho}_{i'j}
\right\}\), respectively  referred to as \(\left\{  \hat{C}_{i'j} \right\}\).
We  also   compute  the  total   cost  of  the  \(j\)-th   path  combination,
\(\tilde{C}_{ij}\) as
\(\tilde{C}_{ij}=\min_{i'\in [U]} \hat{C}_{i'j}\).
After iterating over all  \(\tilde{N}_i=\prod_{i'\in[U] \backslash i} N_{i'}\) combinations, we then find the path combination
\begin{equation*}
\hat{\jmath}=\underset{j'\,\in\,[\tilde{N}_i]}{\mathrm{argmax}}\; \tilde{C}_{ij'}
\end{equation*}
that yields the largest combination cost,
\(C_i=\max_{j\in [\tilde{N}_i]} \tilde{C}_{ij}\).
We  refer to  the  paths  in the  \(t\)-th  tree  of the  \(\hat{\jmath}\)-th
combination as \(\rho^i_t= \hat{\rho}_{i \hat{\jmath}}\).

After repeating the above process for each user,
we compare  the cost  of their  optimal combinations  to determine  the path
with the lowest cost for each user. Let

\begin{equation*}
 \hat{\imath} = \underset{i'\,\in\,[U]}{\mathrm{argmax}}\; C_{i'}
\end{equation*}

be  the user  index with  the largest combination  cost. Then,  the best  path for  the
\(i\)-th user is
\(\rho_i= \rho_i^{\hat{\imath}}\).

\subsection{Determining Path with the Lowest Cost in a Tree}
\label{sec:org6853c8d}
We now  describe  the  algorithm  that
determines  the path  with  the lowest  cost  in an  \(i\)-th  tree, given  a
combination  of  dependent paths  \(\left(  \hat{\rho}_{i'j}\right)_{i'\in[U]
\backslash i}  \in \prod_{i'\in[U] \backslash  i} \symcal{P}_{i'}\)  for the
\(i\)-th user.

Given the \(i\)-th tree, we traverse it  in a depth-first fashion in the same
manner as when we construct the tree, but with a couple of additions.

\begin{itemize}
\item At  each visited  node \(n_i\),  we compute the  link cost  (i.e., SNIR),
\(C_{ij}\), between it and its parent node \(n_j\) according to Eq. (\ref{eq:3}).
\item During  the backtracking  process, at  each node we  decide on  the best
sub-path to any of the leafs. Consider  a node \(n_i\) in the tree that has
a degree  \(d_i\), with \(d_i-1\) children,  \(n_k\), \(k\in[d_{i-1}]\), and
parent  \(n_j\).   Furthermore,  we  know  the  best  sub-paths
\(\bar{\sigma}_k\)  (and its  associated  cost,  \(\bar{C}_{k}\)) from  the
\(k\)-th node to a leaf. Then, the cost at node \(i\) is

\begin{equation*}
\bar{C}_i=\min\left\{ \max_{k\,\in\,[d_i-1]} \bar{C}_k, C_{ij} \right\},
\end{equation*}

and the corresponding optimal sub-path to a leaf is

\begin{equation*}
\bar{\sigma}_i =
        \underset{\bar{\sigma}_k, k\,\in\,[d_i-1]}{\mathrm{argmax}}\;
           \bar{C}_k.
\end{equation*}

It should be noted that for a leaf node \(n_{i'}\) that has node \(n_{j'}\)
as its parent, we have \(\bar{C}_{i'}=C_{i'j'}\), and \(\bar{\sigma}_{i'}\)
as the node itself.
\end{itemize}

The  optimal cost  and the  corresponding path  are determined  when we  have
backtracked back to the root node, after traversing all nodes in the tree.

\subsection{Grouping}
\label{sec:org51f61e3}

Notice  that the  algorithm performs  a tree  search (see  Sec. \ref{sec:org6853c8d})  for each  combination of dependent  paths. Since,
there are  \(\prod_{i'\in[U]  \backslash i}  N_{i'} \)
possible  combinations for each user \(i\),  the search  space  dramatically  increases with  the
number of users, \(U\). Thus,  the proposed algorithm becomes unfeasible when
the system contains a large number of users.

We  introduce the  concept of  grouping that  allows our  algorithm to  scale
nicely  with the  number of  users.  However,  this comes  at the  expense of
the algorithm choosing paths with higher costs. We achieve  this by splitting users into  \(G\) disjoint, equally
sized  groups, and  then  running our  proposed algorithm  on  each of  these
smaller groups. With  this modification, we now have a  total search space of
\(\sum_{j=0}^{G-1}\sum_{i\in[U_j]} \prod_{i'\in[U_j] \backslash  i}
N_{i'}\), which is much smaller than the original algorithm.

\subsection{Complexity Analysis}
\label{sec:complexity-analysis}  There   are  two  parts   to  the
  algorithm: the tree  construction and the path  search. Typically of
  the two, searching  is the more computationally intensive task,  and as such
  we focus on analyzing its complexity.
We assume that the trees pertaining
to  each user  in the  mesh network  is precomputed.   Furthermore, the  trees
contain only valid paths. With the  above assumptions, our search algorithm is
essentially a modified Depth First  Search (DFS) algorithm.

Given these assumptions, the following theorem gives an upper bound on the
time-complexity of our algorithm.
\begin{theorem}
  \label{thm:time-complexity}
  For a mesh network with \(U\) users that are divided into \(G\) groups with
  each group, \(j\in[G]\) having \(U_j\) users, our group-based algorithm has
  a time complexity, \(\Theta\) that can be upper bounded as
  \[
    \Theta\leq\sum_{j\in[G]}\sum_{i\in[U_j]}
    \left(\prod_{i'\in[U_j]\setminus i}
      N_{i'}\right)O(8v+(v+3(U_j-1))x),
  \]
  where \(N_i\)  is the  number of  valid paths  for User  \(i\), and
  \(x\) is  the time complexity for  calculating the SNIR  at a node  in the
  network.
\end{theorem}

\begin{IEEEproof}
  The proof is given in Appendix~\ref{sec:proof-theorem}.
\end{IEEEproof}

\section{Results}
\label{sec:org7779457}


We  consider  a  randomly  generated  mesh network, where the
BSs   are  placed  uniformly   and  randomly  in  a   grid  between
\((0^{\circ},0^{\circ})\) and \((0.01^{\circ},0.01^{\circ})\)  with a specific
constraint that the BSs are at least forty meters apart.  The active users are
also placed in the same grid in an uniform and random fashion, but without the
aforementioned constraint. The users are connected  to the two closest BSs via
an edge, whereas any  two BSs are connected with a  probability of \(0.5\) via
an edge iff  they are within a  minimum distance of \(500\)  meters. We set
the maximum number of hops for a valid path to be \(h_{\mathsf{max}}=4\). We assume
active users and base station have a transmit power, \(P^{\symup{tx}}=30\) dBm
Equivalent Isotropic Radiated Power (EIRP)\footnote{This is a  typical number that has been verified  in lab experiments
  and  agrees  with  EIRP  values  that  can  be  achieved  with  of-the-shelf
  components and antenna designs.},
and transmit at  a frequency \(f=60\) GHz.  Furthermore, all links
  assume a  fading margin of  0.0205 dB/m and  a signal attenuation  factor of
  0.016 dB/m.  All links  between the  base stations  suffer from  a constant
thermal noise. The  BSs are assumed to  use a circular array  antenna, and can
beam-form with an array gain of 20 dB. A sine-based antenna pattern is assumed
to model interference.

\begin{center}
\begin{figure}[t]
\centering
\includegraphics[width=0.9\columnwidth]{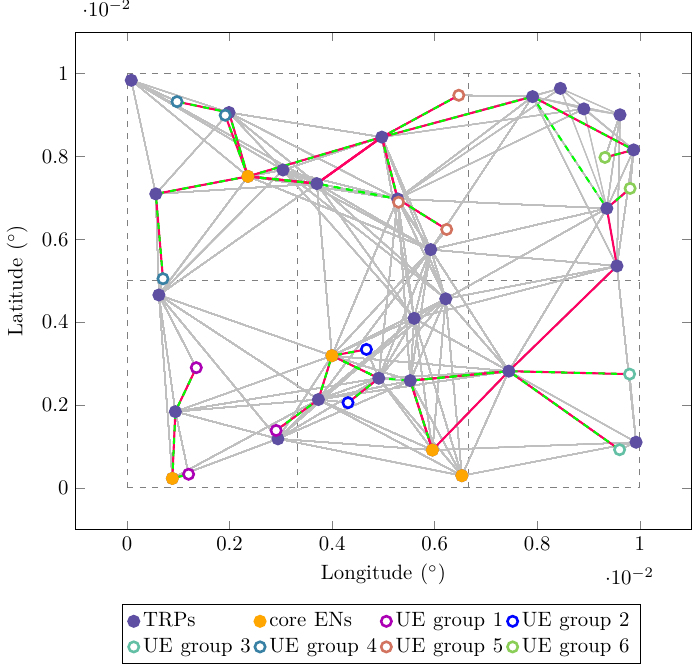}
\caption{\label{fig:big_network}A mesh network representing 30 BSs, of
  which 5 are core BS and 15 users that are divided into 6 groups.   The gray edges in the network represent the links between the BSs, or a BS
  and a user.
The paths
  chosen by our grouping based algorithm  are represented in red color, while
  the paths in green color are chosen by Algorithm B.}
\end{figure}
\end{center}
In Fig.~\ref{fig:big_network}, we illustrate a mesh network that consists of \(30\) BSs,
of which  \(5\) are  connected to  the core network,  and there  exists \(15\)
users in the network that are divided into \(6\) disjoint groups.  We have set
the noise  power to, \(P^{\mathsf{n}}=-100\)  dBm.  We use our  grouping based
algorithm to  find the  optimal paths in  the network. These  are shown  in red
color  in  the  figure. In  contrast,  when  the  noise  power is  set  to  an
arbitrarily high  value, \(P^{\mathsf{n}}=30\) dBm,  i.e., when the  affect of
interference  is negligible  compared to  noise power,  our routing  algorithm
chooses a  different set of  paths. These are highlighted  in green in
the  figure.



\begin{table}[t]
  \centering
  \vskip -1.5em
  \caption{Performance of different algorithms. Each entry in second
    column onwards is the CoA in dB for different mesh networks.}
  \begin{tabular}{lcccccc}
    \hline
    \hline\\[-2ex]
    \hspace{4ex}Mesh & \multirow{2}{2em}{Ours} & \multirow{2}{3em}{Alg. B} & \multirow{2}{3em}{Alg. C} & \multicolumn{3}{c}{GA}\\
    \((B,U,C,G)\) &  &  &  & min & max & avg\\
    \hline\\[-1.5ex]
    \((10, 4, 3, 1)\)  & \(15.63\) & \(-1.63\) & \(-0.18\)  & \(12.73\) & \(15.63\) & \(15.20\)\\[2pt] 
    \((20, 10, 3, 4)\) & \(12.81\) & \(9.70\)  & \(-4.92\)  & \(8.09\)  & \(12.78\) & \(11.14\)\\[2pt] 
    \((30, 15, 5, 6)\) & \(19.00\) & \(0.3\)   & \(-0.899\) & \(7.46\)  & \(14.83\) & \(11.87\)\\[2pt] 
    \hline
    \hline\\[-1.5ex]
  \end{tabular}  
  \label{tab:performance-table}
  \vskip -1.5em
\end{table}

\begin{table*}[t]
  \centering
  \caption{Complexity  comparison  between   our  algorithm\protect\footnotemark
    and 50  runs of the  GA. \(x\) is the  time complexity of  determining the
    SNIR in a node, while \(y\) is the time complexity for determining a valid
    path cost.}
  \vskip -1em
  \begin{tabular}{lcccc}
    \hline
    \hline\\[-2ex]
    \hspace{4ex}Mesh & Avg. hops & \multirow{2}{10em}{\centering Ours (Upper bound)} & \multicolumn{2}{c}{GA}\\
    \((B,U,C,G)\) & required & & exact & avg. \\
    \hline\\[-1.5ex]
    \((10,4,3,1)\)  & 3 & \(O(779560+134428x)\) & \(O(154400+80000y)\) & \(O(314400+160000x)\)\\[2pt]
    \((20,10,3,4)\) & 3 & \(O(529496+75940x)\)  & \(O(1.661\cdot10^6+10^6y)\) & \(O(3.661\cdot10^6+2\cdot10^6x)\)\\[2pt]
    \((30,15,5,6)\) & 3 & \(O(1.52\cdot10^9+192.44\cdot10^6x)\) & \(O(24.244\cdot10^6+3.75\cdot10^6y)\) & \(O(31.744\cdot10^6+7.5\cdot10^6x)\)\\[2pt]
    \hline
    \hline\\[-1.5ex]
  \end{tabular}
  \label{tab:complexity-table}
  \vskip -2em
\end{table*}

{
  In the following, we consider  two algorithms, Algorithms B and C,
  to compare with  our group-based algorithm. Algorithm B is  a variant of our
  algorithm  that   does  not  consider   the  effects  interference   in  the
  network. This  is achieved  by artificially  setting the  noise power  to an
  arbitrarily  high value.  Algorithm C  is  a trivial  random algorithm  that
  randomly chooses  a path for each  from a set  of all possible paths  to the
  core BSs in the network.}
To  quantify  the effectiveness  of  our  algorithm,  let  us define  a  term,
\emph{Cost of Algorithm} (CoA). We define the CoA to be the minima of the path
costs of the optimal path for each user determined by the algorithm. Since the
path costs  of the algorithms is  in dB, the unit  of CoA is also  dB.  We use
this term to quantify the performance of an algorithm because often times in a
network, it is the  worst path that determines the overall  performance.  In
Table~\ref{tab:performance-table} we  tabulate the CoA of our algorithm,
Algorithms B and C,  and  \((K, J)\) GA \cite{madapatha2021topology}. The entries for Algorithm
C are the average  CoA obtained after running it for  1000 times. Each network
in the table is randomly created with
the  same network  parameters as described earlier in this section.  We  choose \(P^{\mathsf{n}}=-100\)  dBm, i.e.,
both noise and  interference power are non-negligible. The GA  is a stochastic
algorithm,  thus we  run it  50  times. For  the three  networks (smallest  to
largest) in the table, the GA runs for \(20\), \(50\), and \(200\) generations
in each run, respectively. Furthermore, the \((K, J)\) parameters used in the
networks are \((20,10)\), \((40,20)\) and \((100,50)\), respectively.

\footnotetext{The  time  complexities  of  our  algorithm  also  includes  the
  complexities  associated to  insignificant  operations  that are  explicitly
  discarded   in    the   proof   of    Theorem~\ref{thm:time-complexity}   in
  Appendix~\ref{sec:proof-theorem}
  .}  We see that our grouping based algorithm
has a CoA that  is at least \(3.11\) dB better than Algorithm  B, and at least
\(15.81\) dB better than Algorithm C.  Moreover, for the parameters chosen for
the GA, we see  that our algorithm outperforms the GA  on an average.  However
the difference  in minuscule for  the \((10,4,3,1)\) network,  suggesting that
the  GA parameters  is  sufficiently large  for such  a  small network,  hence
allowing the  GA to perform  close to  optimality. This indicates  that having
larger GA  parameters will improve the  performance but at the  cost of higher
complexity.

The GA has lower complexity than  our algorithm, but its stochastic nature can
lead       to      inconsistent       performance.      As       shown      in
Table~\ref{tab:performance-table}, the minimum  CoA is at least  2.47 dB worse
than the average, and 2.9 dB worse  than the maximum CoA across 50 runs. Thus,
network   engineers  may   prefer  multiple   GA   runs  to   find  the   best
result. Table~\ref{tab:complexity-table}  compares the total complexity  of 50
GA runs  with a  single run  of our  deterministic algorithm.  Both algorithms
assume  all paths  from the  user to  the  core BSs  are known  via the  trees
described in  Sec.~\ref{sec:org6484d8d}, so the  cost of tree  construction is
excluded.  Complexity is  presented in  terms of  \(x\), the  SNIR computation
complexity,  and \(y\),  the  path cost  complexity. Since  path  cost is  the
maximum SNIR  across nodes, it  depends on \(x\) and  the number of  hops (see
Eq.~(\ref{eq:17})
). We've calculated  the average number of hops across
all valid paths to evaluate the average GA complexity over 50 runs in terms of
\(x\).  The  computation  of  SNIR  involves a  series  of  64-bit  arithmetic
operations, thus we lower bound \(x>64\). Our algorithm shows lower complexity
in  the   \((10,4,3,1)\)  and  \((20,10,3,4)\)  mesh   networks,  with  better
performance than the  GA. Although increasing the GA  parameters could improve
its performance, it  would still be less efficient than  our algorithm. In the
larger \((30,15,5,6)\) network, the GA  has lower complexity but significantly
worse performance, trading off complexity for performance.

\section{Conclusion}
\label{sec:conclusion}

We proposed a  novel tree search-based routing algorithm for  mesh networks at
the edge of  wireless networks.  Our algorithm accounts  for interference from
active communication links and identifies  optimal paths for each user. Across
various  network sizes,  our algorithm  outperforms one  that ignores  network
interference by at least \(3.11\) dB and up to \(18\) dB. It also surpasses an
algorithm  that randomly  selects paths  by at  least \(15.81\)  dB and  up to
\(19.90\) dB.  Furthermore, our algorithm  is at least  \(0.43\) dB and  up to
\(7.13\) dB better than the recently proposed GA.  In scenarios where the GA's
performance  is within  \(2\) dB  of our  algorithm, we  demonstrate that  our
algorithm has lower complexity.

\appendices

\section{Proof of Theorem~\ref{thm:time-complexity}}
\label{sec:proof-theorem}
We begin the proof by outlining the major steps involved in our algorithm.  In
particular,   for  a   given  \(j\)-th   group,  \(j\in[G]\),   and  a   tree,
\(T_i,i\in[U_j]\), we perform  the following steps for each tree  in the group
for all groups.

\begin{enumerate}
\item We  use the DFS algorithm  to determine the  best path with its  cost in
  \(T_i\), assuming a set of paths (we refer to them as dependent paths)---one
  from      each       tree---from      the      remaining       trees      in
  \(T_{i'}, i'\in[U_j]\setminus i\).
\item Given  the best  path in  \(T_i\), we  find the  cost of  each dependent
  paths. For  simplicity we  call this  set of  paths as  a solution  set, and
  determine the cost  of the solution to  be the minimum of all  path costs in
  the solution set.
\item We repeat the two steps for each combination of dependent paths.  
\item Finally, we choose the best solution, which is the one that has the
  maximum solution cost.
\end{enumerate}

After performing the above steps for the \(j\)-th group, the algorithm
computes the optimal paths for each user in the group. This is trivially done
by choosing paths  whose corresponding solution set has the largest solution
cost across the \(U_j\) solution sets  obtained in Item~4.

\IncMargin{1em}
\begin{algorithm}[t]
  \label{alg:pseudo-algorithm}
  \caption{Pseudo algorithm of the modified DFS}
  
  \SetAlgoLined

  \SetKwInOut{Input}{input}
  \SetKwInOut{Output}{output}
  \SetKwFunction{DFS}{DFS}
  \SetKwFunction{argmin}{argmin}
  \SetKwFunction{max}{max}
  
  \Input{Node \(u\) in the tree}
  \Output{Best path from core node to Node \(u\), \(path\) and cost at Node
    \(u\), \(cost\)}
  
  \(node[u]\gets true\)\;
  \(snir\gets\) compute SNIR for Node \(u\)\;
  \For(\tcc*[f]{for each adj. node}){\(n\in\mathcal{N}_u\)}{
    \If{node[n] = false}{
      \(nCost, nPath\gets\DFS{n}\)\;
      append \(nCost\) to \(costList\)\;
      append \(nPath\) to \(pathList\)\;
    }
  }
  \(idx\gets\argmin{costList}\)\;
  \(minCost\gets costList[idx]\)\;
  \(path\gets pathList[idx]\)\;
  \(cost\gets \max{snir, minCost}\)\;
\end{algorithm}

We will now briefly outline the time-complexity of each of the items in the
above enumerated list.

The modified DFS algorithm  that we use in first step above, which has been described
in  previous section,  can  be aptly  be summarized  using  a pseudo  algorithm
described in Algorithm~\Ref{alg:pseudo-algorithm}.The  time complexity of this
algorithm is  simply the sum  of time complexities  of each line.   Looking at
lines 1, 4,  6, 7, 11, 12 and  13, they have \(O(1)\) complexity  each. Line 2
has  a non-trivial  complexity,  and for  now lets  denote  its complexity  by
\(O(x)\).  Assume  that  Node  \(u\)  has degree  \(d\).  Then,  the  variable
\(costList\) has  exactly \(d-1\) elements  that pertain to the  \(d-1\) child
nodes that the algorithm has traversed from  Node \(u\).  We can then say that
Line 10 has \(O(d-1)\)  complexity. Notice that Lines 1, 2, 10,  11, 12 and 13
run once. Furthermore,  after traversing through the whole tree,  Lines 3 to 9
have run approximately the same amount as there are edges in the tree. Let the
number of  nodes in the tree  be equal to  \(v\), then the tree  has \(e=v-1\)
edges. Thus,  the total complexity of the modified DFS algorithm, after  traversing through each node  in the
tree, is

\[  O(4v+vx+e+3e)=O(4v+vx+4v-4).\]

The time complexity of Item 1 is the sum of the complexity of the modified DFS
and the complexity for choosing \(U_j-1\) paths from remaining trees in the
set \(\{[U_j]\setminus i\}\). Since the complexity for choosing the remaining
paths is \(O(U_j-1)\), the total complexity of Item~1 is
\begin{equation}
  \label{eq:13}
  O(4v+vx+4v-4+U_j)\approx O(8v+vx).
\end{equation}

Assume the time complexity for finding cost of each path to be \(O(y)\). Then,
the complexity of Item 2 is simply

\begin{equation}
  \label{eq:14}
  (U_j-1)O(y)+O(U_j)=O((U_j-1)y+U_j)\approx O((U_j-1)y),
\end{equation}
where the first  term in the leftmost expression comes  because we compute the
cost of  \(U_j-1\) dependent paths, and  the last term is  time complexity for
finding the minimum from list of \(U_j\) path costs.

Adding the  complexities in  Eq.~(\ref{eq:13}) and Eq.~(\ref{eq:14}),  and multiplying
with    the    number    of    dependent   path    combinations    gives    us
the time complexity of Item 3. This amounts to
\begin{equation}
  \label{eq:20}
  \left(\prod_{i'\in[U_j]\setminus  i}N_{i'}\right)O(8v+vx+(U_j-1)y),
\end{equation}
where \(N_{i'}\) is the number of valid paths in \(T_{i'}\).

In Item  4, one performs an  argmax operation to determine  the solution index
that has the  highest cost, and then obtain the  corresponding solution from a
list    of     solutions.     The    former    has     the    complexity    of
\(O\left(\prod_{i'\in[U_j]\setminus  i}N_{i'}\right)\), while  the latter  has
\(O(1)\). Since the latter is much smaller than the former, one can
approximate the time complexity of Item 4 as

\begin{equation}
  \label{eq:19}
  O\left(\prod_{i'\in[U_j]\setminus i}N_{i'}\right)
\end{equation}

Finally, the overall complexity of four steps is obtained by adding
Eq.~(\ref{eq:20}) and Eq.~(\ref{eq:19})

\begin{equation}
  \label{eq:15}
  \begin{split}
    \left(\prod_{i'\in[U_j]\setminus
    i}N_{i'}\right)O(8v+vx+(U_j-1)y+1)\\
    \approx
    \left(\prod_{i'\in[U_j]\setminus i}N_{i'}\right)O(8v+vx+(U_j-1)y)
  \end{split}
\end{equation}

As  described earlier,  the last  step of  the algorithm  is to  determine the
optimal paths for the users within a group. This requires finding the solution
set with  maximum solution cost  amongst \(U_j\) solution sets,  and obtaining
the paths for  the corresponding solution set.  Doing an  argmax operation has
the  complexity of  \(O(U_j)\), and  obtaining the  set of  paths from  a list
containing set of  paths for each corresponding solution set  has a complexity
of  \(O(1)\).    Notice  that  these   complexities  are  much   smaller  than
Eq.~(\ref{eq:15}), and thus  the overall complexity per group  can be approximated
as

\begin{equation}
  \label{eq:16}
  \Theta_j=\sum_{i\in[U_j]}
  \left(\prod_{i'\in[U_j]\setminus i}N_{i'}\right)O(8v+vx+(U_j-1)y)
\end{equation}

The time-complexity of finding  the cost of a valid path  can be upper bounded
as
\begin{equation}
  \label{eq:17}
  y\leq 3O(x)+O(3).
\end{equation}
This is because the  core node in each valid path is at  most 4 hops away from
the user. Thus, at the most, we  compute SNIR for 3 nodes (nodes excluding the
user and the first base station that it connects to), and then assign the path
with a cost that  is the minimum of the three. The former  has a complexity of
\(O(x)\), while the latter has the complexity of \(O(3)\).

Substituting Eq.~(\ref{eq:17}) in Eq.~(\ref{eq:16}) we get an upper bound on the
time-complexity of our algorithm as
\begin{equation}
  \label{eq:18}
  \Theta_j\leq\sum_{i\in[U_j]}
  \left(\prod_{i'\in[U_j]\setminus i}N_{i'}\right)O(8v+(v+3(U_j-1))x),
\end{equation}
where, as  mentioned previously, \(x\)  is the complexity for  determining the
SNIR at  a node.  Finally, summing  Eq.~(\ref{eq:18}) for  all groups,  proves the
theorem.
  
\bibliographystyle{ieeetr}
\bibliography{hexaX.bib}

\end{document}